\documentclass[a4paper]{jpconf}
\usepackage{graphicx}
\begin{document}
\title{Low Energy Investigations at Kamioka Observatory}

\author{Hiroyuki Sekiya}

\address{Kamioka Observatory, Institute for Cosmic Ray Research, University of Tokyo, 456 Higashi-Mozumi, Kamioka, Hida, Gifu, 506-1205 Japan}

\ead{sekiya@icrr.u-tokyo.ac.jp}

\begin{abstract}
At Kamioka Observatory many activities for low energy rare event search are ongoing.
Super-Kamiokande(SK), the largest water Cherenkov neutrino detector, currently continues data 
taking as the fourth phase of the experiment (SK-IV). In SK-IV, we have upgraded the water 
purification system and tuned water flow in the SK tank. Consequently the background level was 
lowered significantly. This allowed SK-IV to derive solar neutrino results down to 3.5~MeV 
energy region. With these data, neutrino oscillation parameters are updated from global fit; 
$\Delta m^2_{12}=7.44^{+0.2}_{-0.19}\times10^{-5} {\rm eV}^2$, 
$\sin^2\theta_{12}=0.304\pm0.013$, 
$\sin^2\theta_{13}=0.030^{+0.017}_{-0.015}$.
NEWAGE, the directional sensitive dark matter search experiment, is currently operated 
as ``NEWAGE-0.3a'' which is a $0.20\times0.25\times0.31$~m$^3$ micro-TPC filled with CF4 gas at 152~Torr.
Recently we have developed ``NEWAGE-0.3b''. It was succeeded to lower the operation pressure down to 76~Torr and
the threshold down to 50~keV (F recoils).
XMASS experiment is looking for scintillation signals from dark matter interaction 
in 1~ton of liquid xenon. It was designed utilizing its self-shielding capability with 
fiducial volume confinement. However, we could lower the analysis threshold down to 0.3~keVee
using whole volume of the detector. In February 2012, low threshold and very large exposure 
data (5591~kg$\cdot$days) were collected. With these data, we have excluded some part of the 
parameter spaces claimed by DAMA/LIBRA and CoGeNT experiments.
%%  We also searched for solar axions which would be produced through bremsstrahlung and Compton effect in the sun. 
%%The resulting limit on the coupling is $g_{aee} < 5.5\times10^{-11}$ which is factor four stronger than a recent experimental limit.

\end{abstract}

\section{Kamioka Observatory}
\begin{figure}
\begin{center}
\includegraphics[width=10cm]{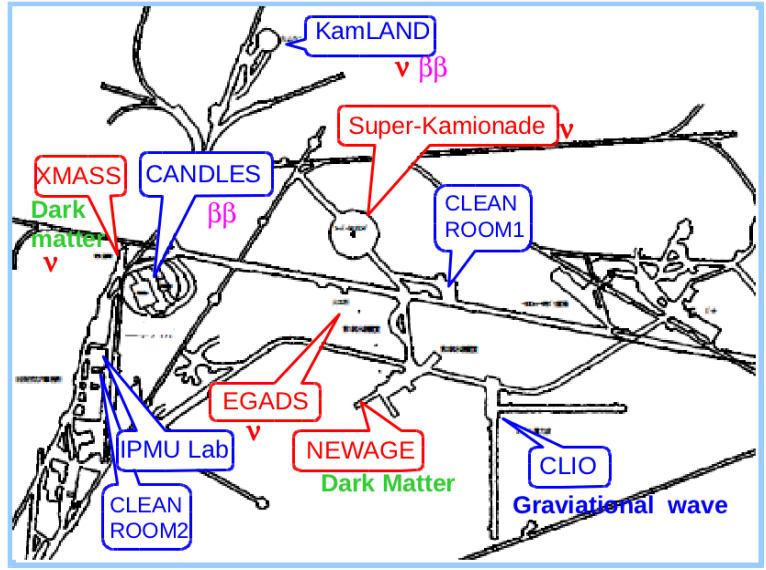}
\caption{
Kamioka underground map of experiments.
}
\label{map}
\end{center}
\end{figure}

Kamioka Observatory, Institute for Cosmic Ray Research, University of Tokyo, is located in Kamioka-town in the northern part of the Gifu prefecture in Japan.
The location of the laboratories is under the peak of Mt. Ikenoyama providing 2,700 meters water equivalent (or 1000~m of rock) overburden.
This observatory was established in 1995 to push forward with the Super-Kamiokande experiment and has become a world frontier center of neutrino physics.
Many studies complementary to the flagship SK are also ongoing as illustrated in Figure~\ref{map}. 
Experiments to search for dark matter and double beta decay are very active now. Further a new 3~km gravitational wave telescope is being constructed and 
the next generation flagship experiment: Hyper-Kamiokande is seriously discussed in Japanese physics community.

\section{Super-Kamiokande}
Super-Kamiokande (SK) is a water Cherenkov detector containing 50,000 tons of pure water viewed by 11,129 
(inner detector:\,32\,kton) and 1,885 (outer detector:\,18\,kton) photomultiplier tubes (PMTs). 
%%Since the upgrade of its electronics in August 2008,
%%SK continues data taking as the fourth phase of  the experiment (SK-IV), and high quality data has been 
%%accumulated.
A broad range of energies is studied with the SK detector, from a few MeV up to $O$(TeV).
In the low energy region below $\sim$20\,MeV, solar neutrino interactions are detected by neutrino-electron
scattering. At higher energies, atmospheric neutrino interactions covering 5 orders of
magnitude are detectable by neutral- and charged-current neutrino-nucleon interactions.

In this presentation, we focus on the lower energy physics, solar neutrino.
SK detects solar neutrinos through neutrino-electron elastic scattering, 
where the energy, direction, and time
of the recoil electron are measured.
Due to its large (22.5\,kton) fiducial mass SK makes 
precise measurements of $^8$B solar neutrinos,
including precision information on their energy 
spectrum and its time variation.

Since SK-III started, we have made many efforts to reduce backgrounds.
%% and to increase the precision of the detector 
%%calibrations, both of which are crucial for solar neutrino measurements.
The most serious background comes from the beta decay of $^{214}$Bi, which 
is produced in the decays of radon in the air and detector materials 
as well as from radium in the water.
In order to reduce the $^{214}$Bi background, the SK water system has been 
upgraded. First, the cooling power of the heat exchanger for the supply water
was increased so as not to induce convection in the tank, which transports
radon near the PMTs into the fiducial volume.
Second, new membrane degasifier units were added to increase the efficiency
of radon removal from the supply water.
Third, the water flow in the detector was precisely investigated
and optimized to reduce the background contamination in the fiducial volume
as much as possible. During the SK-IV period we have introduced a precise
temperature control system for the inlet water to further reduce
convection in the water.

As a result of these improvements, the background rate in the lower energy
region in SK-IV has been reduced by a factor of more than three compared 
to the SK-I period. Further, it has allowed a lower analysis energy threshold.
%%down to 3.5~MeV of kinetic energy.
Until the end of March 2012, 1069.3 days of SK-IV solar neutrino data for
analysis was taken. A clear solar neutrino signal in the 3.5-4.0\,MeV
kinetic energy region was observed at more than $7\,\sigma$.
In addition, we developed a new analysis method for these low energy regions
based on the amount of multiple Coulomb scattering of electrons using the PMT hit
pattern of the Cherenkov cone. Lower energy electron hit patterns 
(i.e. the $^{214}$Bi background) are more isotropic than those of the higher
energy solar neutrino recoil electrons.
 This analysis improves the statistical
uncertainty of the number of signal events by about 10\,\%. 
We use this new method for recoil energies below 7.5\,MeV.

%%as shown in
%%Figure.\ref{sk:sol01}. 
%%\begin{figure}
%%\begin{center}
%%\includegraphics[width=5cm]{sk-sol01.pdf}
%%\caption{
%%Solar angle distributions of 1069.3 days SK-IV data sample with energy between 3.5-4.0\,MeV.
%%}
%%\label{sk:sol01}
%%\end{center}
%%\end{figure}

%%The systematic uncertainty on the total flux in the energy region between 4.0 and 19.5 MeV
%5during SK-IV becomes $\pm1.7\%$, which is nearly half that 
%%of the SK-I period, $^{+3.5}_{-3.2}\%$~\cite{sk:solar_sk1full}.
%%This reduction comes from improvements in the uncertainty of the
%%fiducial volume size ($\pm$1.3\% in SK-I, $\pm$0.17\% in SK-IV), 
%%better understanding of the absolute energy scale ($\pm$0.64\% in SK-I, 
%%$\pm$0.54\% in SK-IV), and careful studies of the data reduction.
%%During SK-IV the measured $^{8}$B flux is \( 2.34\pm0.03(stat.)\pm0.04(sys.)\times
%%10^{6}cm^{-2}s^{-1} \) , which is consistent with previous measurements from SK-I, II,
%%and III and is shown in Figure.\ref{sk:sol02}. 

\begin{figure}
\begin{center}
\includegraphics[width=10cm]{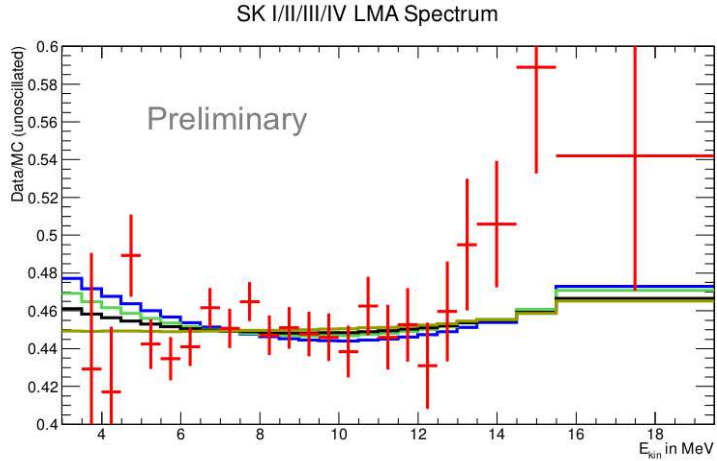}
\caption{
 SK-I to SK-IV combined solar neutrino energy spectrum.
Each point shows the ratio of the data to the expected flux 
using an unoscillated $^8$B solar neutrino spectrum.
Predictions for (1) $\sin^2\theta_{12}=0.304$ and 
$\Delta m_{21}^2=7.4\times10^{-5}$eV$^2$(blue),
(2) $\sin^2\theta_{12}=0.314$ and 
$\Delta m_{21}^2=4.8\times10^{-5}$eV$^2$(light blue),
(3) flat probability (black), and (4) flat probability and
$d\sigma / dE$ shape for pure $\nu_e + e$ scattering (gold)
are also shown.
}
\label{sk:sol03}
\end{center}
\end{figure}

%%The measured energy spectrum from SK-IV is shown in Figure.\ref{sk:sol03}.
Figure~\ref{sk:sol03} shows SK-I to SK-IV combined energy spectrum
with expectations from the solar global and solar+KamLAND as well as
flat reduction of the neutrino spectrum.
The vertical axis shows the ratio of the observed energy spectrum to the 
expectation from the unoscillated MC simulation assuming a  $^8$B flux of 5.25$\times 10^6$\,cm$^{-2}$s$^{-1}$.
The combined energy spectrum is consistent with the flat prediction, 
but the level of favoring flat over the upturn is 1.1$\sim$1.9$\sigma$ level.

\begin{figure}
\begin{center}
\includegraphics[width=10cm]{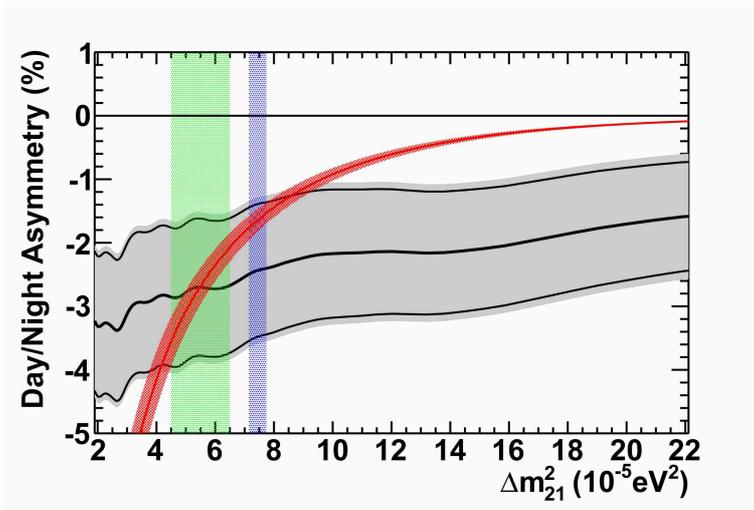}
\caption{
The day-night asymmetry as a function of $\Delta m^2$.
A thick black line shows the result of the SK fit and the surrounding gray
band indicates statistical and systematic uncertainty (thin black lines for
statistical error only).
The red curve shows the expectation assuming $\sin^2(\theta_{12})=0.314$.
The light blue and light green vertical bands show the allowed 
range of $\Delta m^2$
from the KamLAND reactor and solar global, respectively.
}
\label{sk:sol05}
\end{center}
\end{figure}

Concerning differences in the day and night fluxes 
the expected flux asymmetry, defined as 
$A_{DN} = (day-night)/\frac{1}{2}(day+night)$, 
is about $-2\%$ based on current understanding of 
neutrino oscillation parameters.
Although this is not a large effect, long term observations by SK
enable discussion of a finite value of the day-night asymmetry.
The $A_{DN}$ value using the combined SK-I to SK-IV data
is $-2.8\pm1.1\pm0.5\%$, which is a 2.3$\sigma$ difference from zero.
Figure.\ref{sk:sol05} shows $A_{DN}$ as a function of $\Delta m^2$ together
with the expectation. 
The observed $A_{DN}$ is consistent with the expectation using
the best fit $\Delta m^2$ from both KamLAND and the global solar analysis.

%%\begin{figure}
%%\begin{center}
%%\includegraphics[width=8cm]{sk-sol04.pdf}
%%\caption{
%%The day-night asymmetry as a function of energy in the combined
%%SK-I,II,III and IV data. The red line shows the predicted amplitude assuming 
%%neutrino oscillations with 
%%$sin^2\theta_{12}=0.314$ and $\Delta m_{21}^2=4.8\times10^{-5}$eV$^2$.
%%}
%%\label{sk:sol04}
%%\end{center}
%%\end{figure}

We performed a global solar neutrino oscillation analysis 
including all SK data
(SK-I\,\cite{sk:solar_sk1full},SK-II\,\cite{sk:solar_sk2full}, 
SK-III\,\cite{sk:solar_sk3full}, and SK-IV) 
as well as the most recent results from 
SNO\,\cite{sno},
the radiochemical experiments\,\cite{radiochemical1,radiochemical2}
and the latest $^7$Be flux measurement from 
Borexino\,\cite{borexino}.
This analysis was then compared and combined with the reactor neutrino results
from KamLAND\,\cite{kamland}.
%%The green contours in Figure.\ref{sk:sol06} show the allowed region of the neutrino oscillation 
%%parameters sin$^{2}\theta_{12}$ and sin$^{2}\theta_{13}$ in 1-$\sigma$ steps
%%from 1 to 5~$\sigma$. 
The obtained range of sin$^{2}\theta_{13}$ from the solar global
analysis is $0.014^{+0.027}_{-0.021}$, while the value of the KamLAND reactor analysis is
$0.031^{+0.038}_{-0.036}$.
%%\begin{figure}
%%\begin{center}
%%\includegraphics[width=6cm]{sk-sol06.pdf}
%%\vspace{-3.5cm}
%%\caption{
%%  Allowed regions of the neutrino oscillation parameters  
%%  sin$^{2}\theta_{12}$ and sin$^{2}\theta_{13}$
%%  from the global solar neutrino analysis (green) and the KamLAND reactor neutrino data (light blue). 
%%  The purple region shows the contour from the combined global solar and KamLAND reactor analyses.
%%  Curves are drawn for each 1-$\sigma$ step between 1 and 5~$\sigma$ for the global solar analysis,
%%  and 1-3$\sigma$ for the KamLAND and solar+KamLAND regions. Contours at 3~$\sigma$ are filled with
%%  their colors.
%%  }
%%\label{sk:sol06}
%%\end{center}
%%\end{figure}
In the combined fit sin$^{2}\theta_{13}$ was found to be 
$0.030^{+0.017}_{-0.015}$, which is a roughly  2~$\sigma$ hint 
that $\theta_{13}$ is different from zero.
This has been already discussed prior to 2010\,\cite{sk:solar_sk3full}.
After 2011, 
the T2K and reactor experiments
%%Double Chooz, Daya Bay, and Reno experiments 
presented indications and later evidence for a finite $\theta_{13}$. 
The combination of their measurements yields  
$\mbox{sin}^{2} \theta_{13} = 0.025^{+0.003}_{-0.004}$
 and the result of the combined analysis of the global solar and KamLAND 
reactor data is consistent with this value.
%%Figureure \ref{sk:sol07} shows the allowed region of neutrino oscillation 
%%parameters in the $\Delta m^{2}_{21}$ and sin$^{2}\theta_{12}$
%%plane assuming $\sin^2\theta_{13}$ is fixed at 0.025.
%%\begin{figure}
%%\begin{center}
%%\includegraphics[width=6cm]{sk-sol07.pdf}
%%\vspace{-2.5cm}
%%\caption{
%%Allowed regions of neutrino oscillation parameters in the $\Delta m^{2}_{21}$ and sin$^{2}\theta_{12}$ 
%%plane with $\sin^2\theta_{13}$ fixed at 0.025
%%from the global solar neutrino analysis (green) and the KamLAND reactor neutrino data (light blue). 
%%The purple area shows the combined contour of the global solar and the KamLAND reactor analyses.
%%The curves are drawn for each 1-$\sigma$ step from 1-5~$\sigma$ for the global solar, and
%%from 1-3~$\sigma$ for the KamLAND and solar+KamLAND results. Contours at 3~$\sigma$ are filled with
%%their colors.}
%%\label{sk:sol07}
%%\end{center}
%%\end{figure}

Assuming $\sin^2\theta_{13}$ is fixed at 0.025,
the obtained parameters from the global solar analysis are
$\Delta m^{2}_{21}=(4.86^{+1.44}_{-0.52})\times 10^{-5}$eV$^{2}$ and
sin$^{2}\theta_{12} = 0.310^{+0.014}_{-0.015}$.
Comparing these values with those from KamLAND, 
($\Delta m^{2}_{21}=(7.49^{+0.20}_{-0.19})\times 10^{-5}$eV$^{2}$ and
sin$^{2}\theta_{12} =0.309^{+0.039}_{-0.029}$), there is a 1.8~$\sigma$ level tension 
in the $\Delta m^{2}_{21}$ results.
Combining the global solar data with KamLAND, the oscillation parameters become
$\Delta m^{2}_{21}=(7.44^{+0.20}_{-0.19})\times 10^{-5}$eV$^{2}$ and
sin$^{2}\theta_{12} = 0.304\pm0.013$.

\section{NEWAGE}
Among the most plausible dark matter candidates, Weakly Interacting Massive Particles (WIMPs),
which are expected to couple to ordinary matter primarily through the weak force, can 
be detected directly through observation of nuclear recoils produced 
in their elastic scattering interactions with detector nuclei~\cite{jungman}.
The most convincing signature of the WIMPs appears in the directions of nuclear recoils. 
It is provided by the earth's velocity through the galactic halo ($\sim$230~km/s) and the distribution
of the nuclear recoil direction shows a large asymmetry.
Hence, detectors sensitive to the direction of the recoil nucleus would have a great potential to identify WIMPs.
Time Projection Chambers (TPCs) with fine spacial resolutions are among
such devices, and we are developing a micro-TPC~\cite{nishimura, sekiya}, which can detect 
three-dimensional fine tracks of charged particles.
Since the energy deposits of WIMPs to nuclei are only a few tens of keV
and the range of nuclei is limited, the micro-TPC should be operated 
at low pressures. We are interested in operating the micro-TPC with CF$_4$\,
because $^{19}$F has a special sensitivity to SD interactions for its unique spin structure.

\begin{figure}
\begin{center}
\includegraphics[height=5cm]{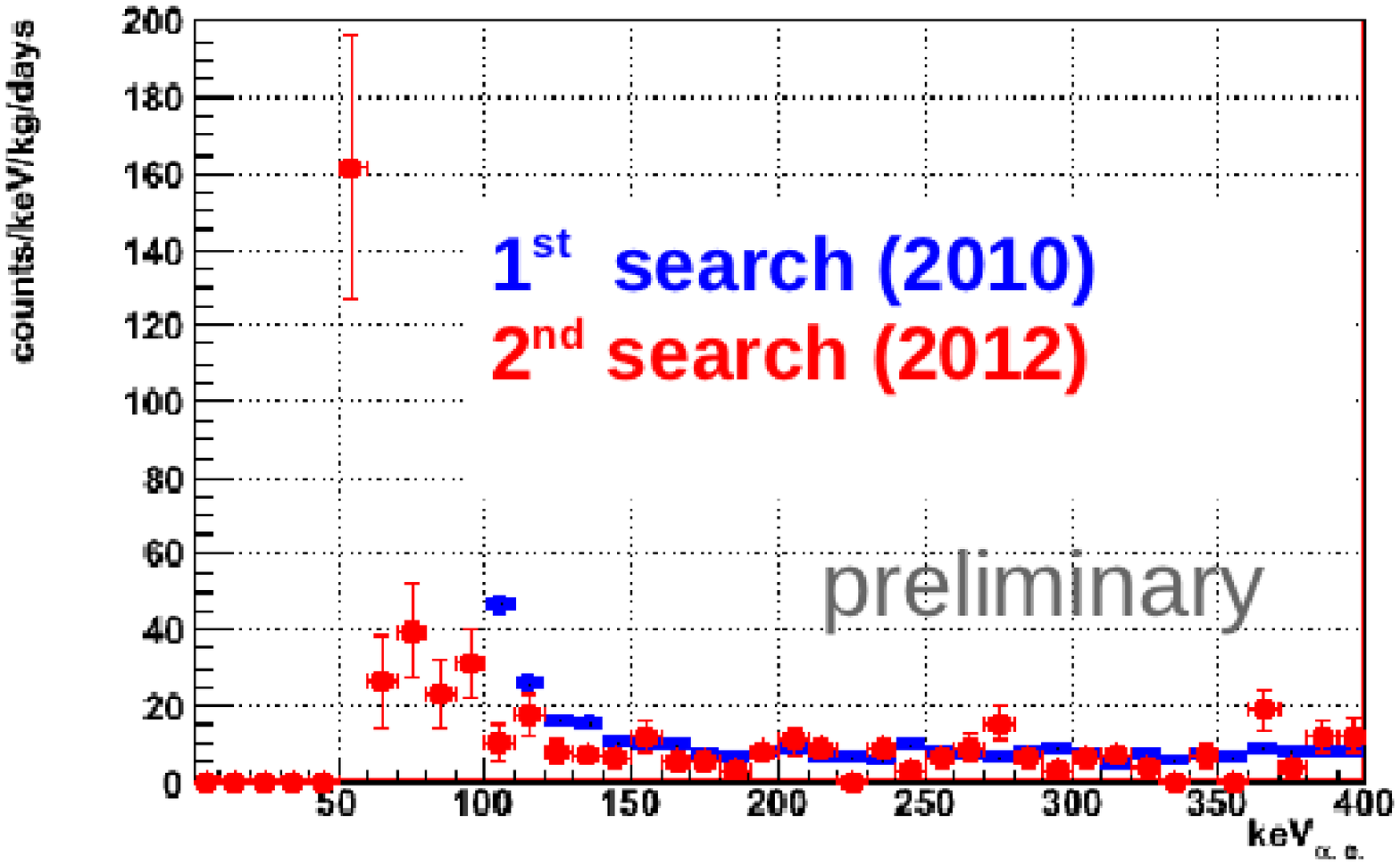}
\includegraphics[height=5cm]{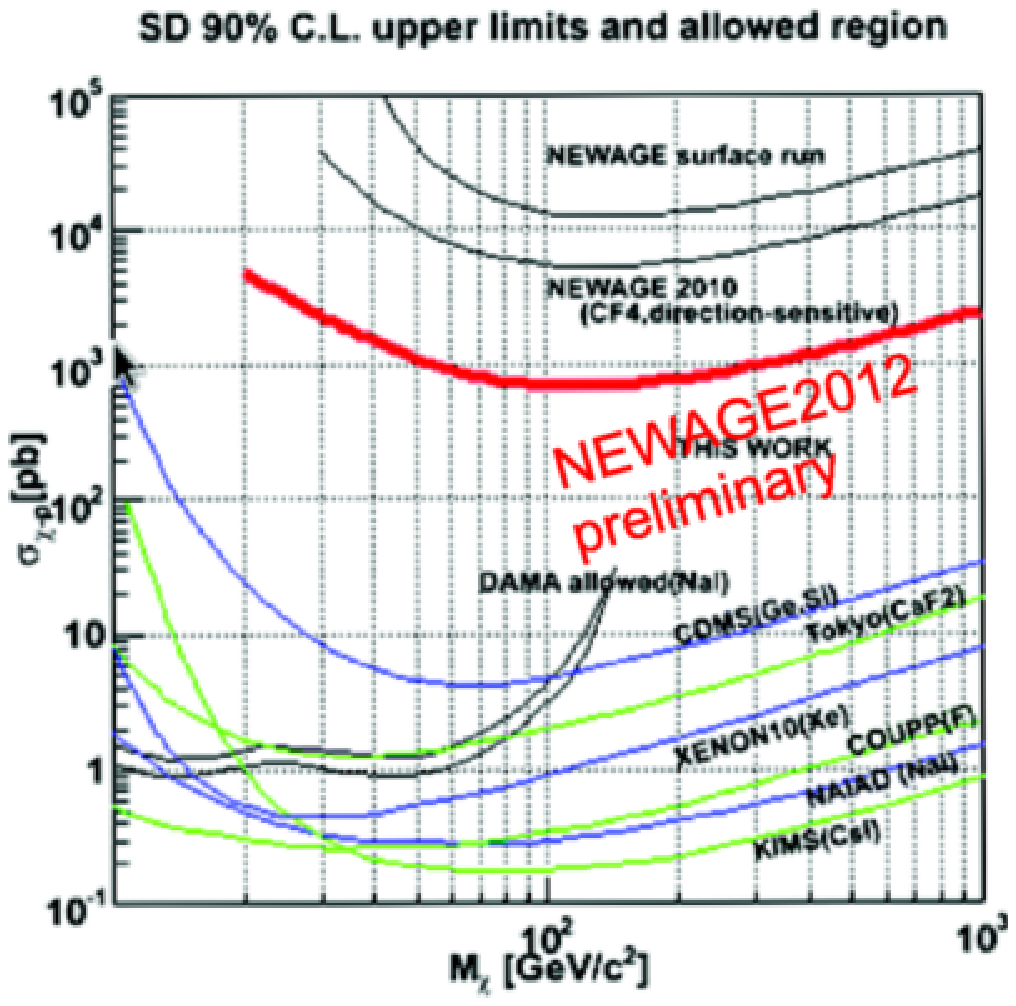}
\caption{
Obtained energy spectra and exclusion limits by NEWAGE-0.3a detector at Kamioka.
}
\label{newlimit}
\end{center}
\end{figure}

We performed our first dark matter search in 2010 at Kamioka laboratory using NEWAGE-0.3a detector~\cite{miuchi}. The exposure was 0.524\,kg$\cdot$days. 
The gas pressure was 152\,Torr and the energy threshold was 100\,keV. Then, second search was conducted from January 2012 to May 2012.
This exposure was 0.140\,kg$\cdot$days with the same gas pressure as before and the energy threshold was lowered to 50\,keV.
Figure~\ref{newlimit} shows the energy spectra and exclusion limits obtained through these searches. We set a limit of
800~pb for 100~GeV WIMPs, however it is indispensable to lower the threshold and background.
In this presentation, we report on the study of lowering the energy threshold via operating the detector in lower pressure.
%%and effors for lowering background.  Rn ni tuite iwanakute yoika mou!!!

\begin{figure}
\begin{center}
\includegraphics[width=8cm]{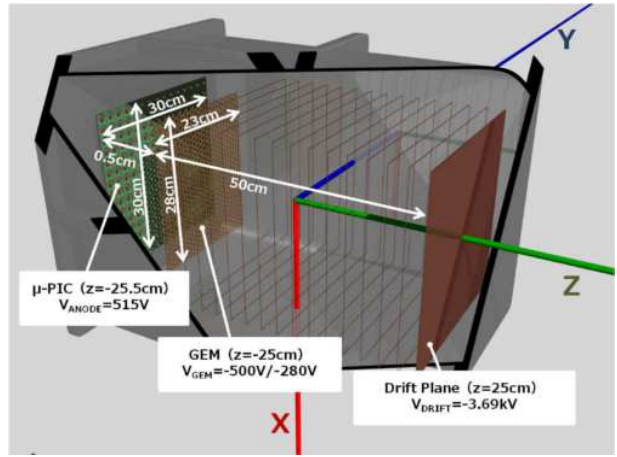}
\caption{
Schematic drawing of NEWAGE-0.3b detector. Induction gap, the distance between $\mu$-PIC and GEM, is 5\,mm.
}
\label{newdet}
\end{center}
\end{figure}

The low pressure gas study was performed using NEWAGE-0.3b detector in the surface laboratory.
A schematic drawing of NEWAGE-0.3b is shown in Figure~\ref{newdet} The size of the $\mu$-PIC
(No.SN070309-3) is $30$\,cm$\times 30$\,cm, and the number of pixels is 768$\times$768. We supply a positive high voltage
to the anode electrodes of the $\mu$-PIC ($= V_{\rm ANODE}$). A large area GEM, consisting of
a 50\,mm-thick polyimide layer sandwiched between 5\,mm-thick copper electrodes is set at 5\,mm
above the $\mu$-PIC to obtain sufficient gain. The size of the GEM is 23\,cm$\times$28\,cm, and this GEM
is segmented into 8 parts to reduce the risk by discharges. We supply negative high voltages at
the top (gas-volume side: $V_{\rm GEM−top}$) and bottom ($\mu$-PIC side: $V_{\rm GEM−bottom}$) of the GEM. 
The drift length is 50\,cm and the detection volume is surrounded by a drift cage to form the electric field. 
The size of the drift cage is $36.3\times36.3\times51$\,cm. Copper electrodes are formed on the inner surface
of the drift cage by every 1\,cm. We supply a positive high voltage ($=V_{\rm DRIFT}$) to the drift plane. 
These detector components are set in a chamber made of SUS304 stainless steel whose inner surface was 
electro-polished to reduce out-gassing. The chamber was filled with CF$_4$ gas at 76\,Torr. 
A dedicated electronics system was used for the measurement. Details of the electronics
are described in~\cite{miuchi} and the references therein. Three-dimensional tracks were detected by
the discriminated ”hit” signals of 768 anode strips and 768 cathode strips and the anode-cathode
coincidence-timing. Energy of each track was determined by the summed waveform.

\begin{figure}
\begin{center}
\includegraphics[width=5cm, angle=270]{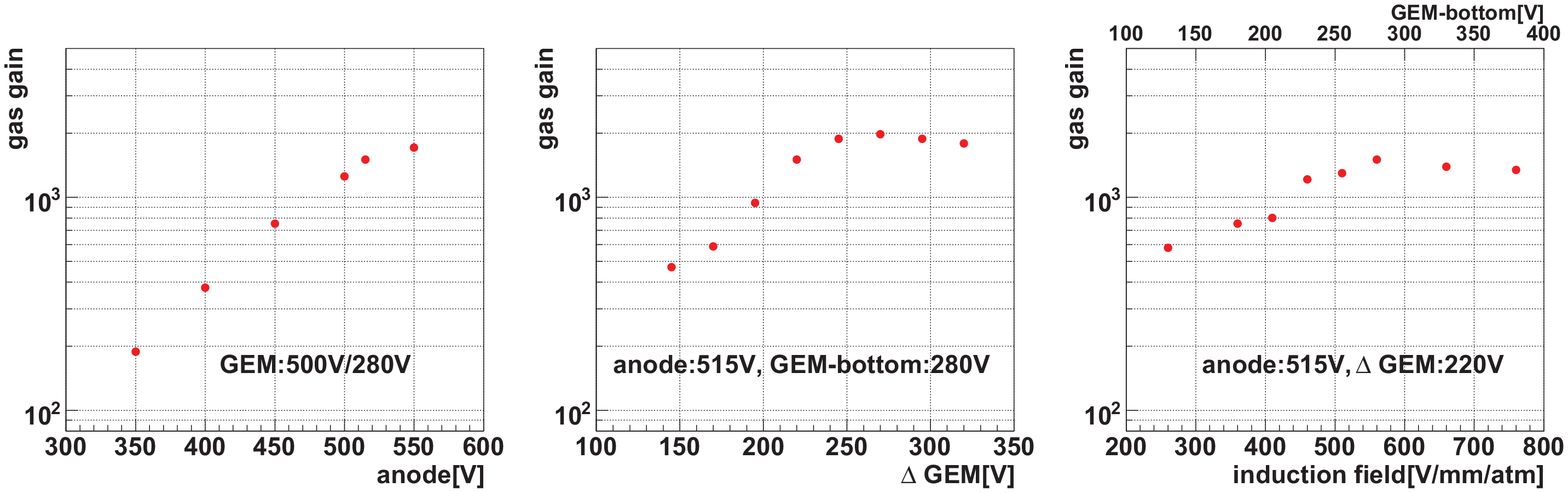}
\caption{
Measured gas gain curves against $V_{\rm ANODE}$, $\Delta V_{\rm GEM}$ and $V_{\rm INDUCTION}$ (or $V_{\rm GEMbottom})$.}
\label{gain}
\end{center}
\end{figure}

\begin{figure}
\begin{center}
\includegraphics[width=5cm, angle=270]{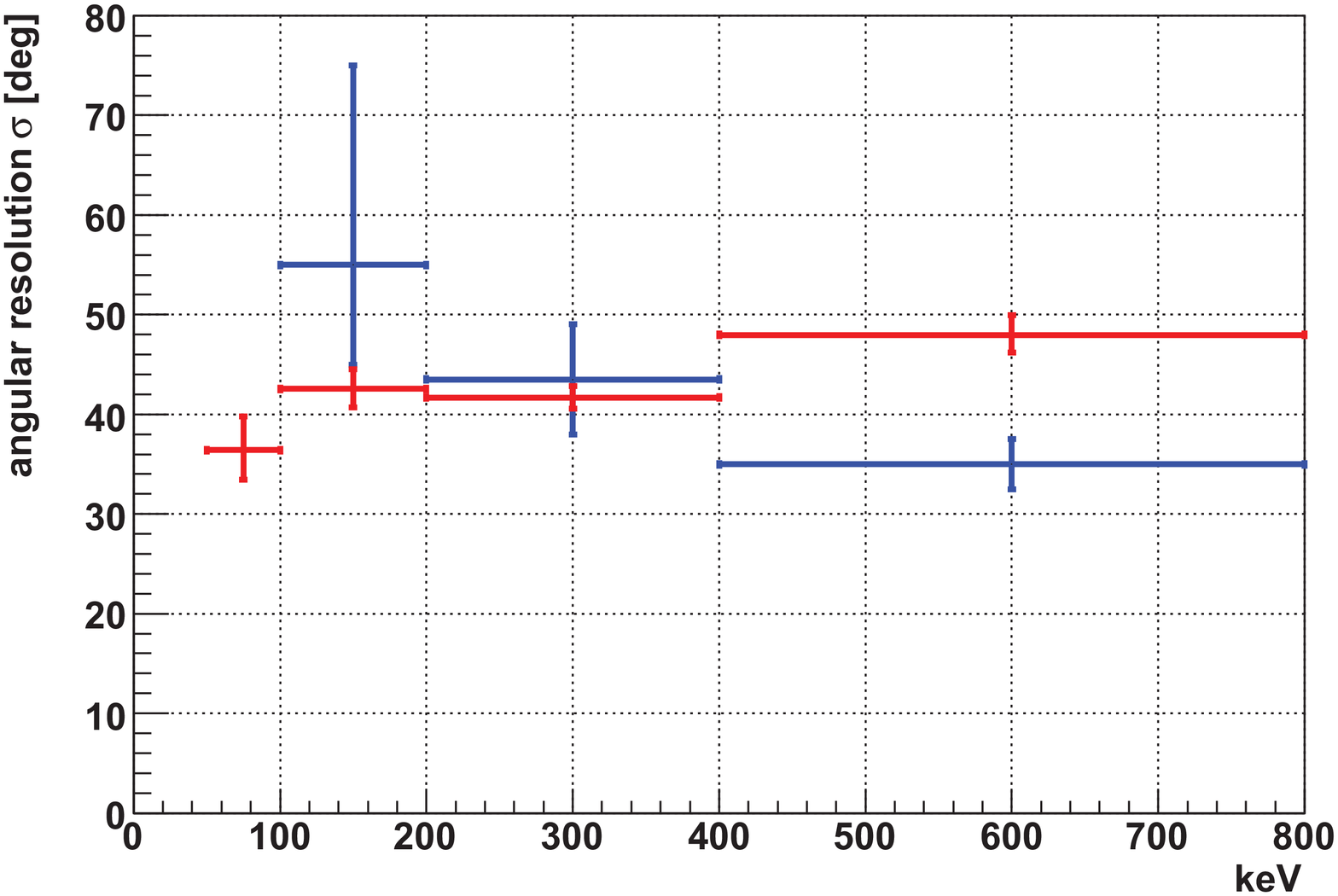}
\caption{
Energy dependence of measured angular resolutions. Red and blue histograms show the results of
76\,Torr and those of 152\,Torr, respectively.
}
\label{angle}
\end{center}
\end{figure}

As a beginning, the gain dependence on the bias voltages was measured to optimize operation parameters
for CF$_{4}$ at 76\,Torr. We had operated the previous NEWAGE-0.3a at a gas gain of 700 with 152\,Torr gas.
Accordingly, the optimum gas gain should be 1400 at 76\,Torr, because the primary electron density became 
1/2 of the previous detector. In Figure~\ref{gain}, the obtained gas gain against $V_{\rm ANODE}$, $\Delta V_{\rm GEM}=V_{\rm GEMtop}-V_{\rm GEMbottom}$, and
the induction field ($V_{\rm INDUCTION}$). The other two parameters are fixed in each plot.
The gas gain increased with $V_{\rm ANODE}$ until the voltage was limited by discharged. 
As for $\Delta V_{\rm GEM}$ and $V_{\rm INDUCTION}$, gain curves saturate below 2000. 
This gain suppression is due to the operation in low gas pressure. The mean free path of an electron 
in 76\,Torr of CF$_{4}$ gas is several mm, and this length is close to the thickness of the GEM (50\,mm),
therefore the gas amplification at the GEM is to be suppressed. Although it turned out that the gain is saturated, 
we have achieved the required gas gain of 1400. The optimized parameters are 
$V_{\rm ANODE} = 515$\,V, $V_{\rm GEM−top }= −500$\,V, $V_{\rm GEM−bottom} = −280$\,V and $V_{\rm DRIFT} = −3.69$\,kV. 
Then, the angular resolution of NEWAGE-0.3b was evaluated using nuclear tracks by neutrons from a $^{252}$Cf source. 
We determined the angular resolution by comparison of measured and simulated distributions of the recoil angle.
The results are shown in Figure~\ref{angle}.

We plan to move the NEWAGE-0.3b detector to Kamioka underground in 2013. 
In order to operate it with lower pressure gas, we need an even higher gas gain. 
As shown in above the gain is saturated at around 2000 in the current set up, 
therefore we have started to test a new GEM which has a thickness of 100\,mm instead of 50\,mm. 
We are also improving the data acquisition system (DAQ) for recording the time-over-threshold (TOT) 
of the $\mu$-PIC signals in order that we can measure the energy deposition at each strip. 

\section{XMASS}
Over the past 10 years, some experiments indicate a possible WIMP signal~\cite{dama,cogent,cresst} 
with a lighter mass of $\sim 10$\,$\mbox{GeV}$ and with a spin-independent cross section of the order of $\sim 10^{-40} \mbox{cm}^{2}$.
These positive signals have come predominantly from experiments without the capability 
to discriminate between electromagnetic and nuclear recoils, while other experiments that have this ability
have excluded light WIMPs at these cross-sections~\cite{xenon,cdms,edelweiss}. 
In this presentation, a search for light WIMPs using a 5591.4\,kg$\cdot$day exposure of the 
XMASS experiment without nuclear recoil discrimination is presented.

\begin{figure}
\begin{center}
\includegraphics[width=11.0cm]{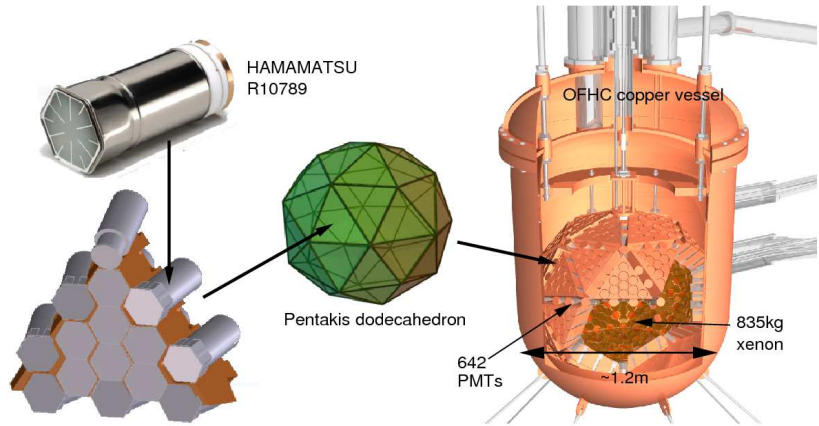}
\caption{Structure of the XMASS detector.}
\label{xmass_detector}
\end{center}
\end{figure}

XMASS is a single phase liquid xenon scintillator detector containing 1050\,kg of Xe in an OFHC copper vessel. 
As shown in Figure.~\ref{xmass_detector} xenon scintillation light is collected by 630 hexagonal and 12 cylindrical 
inward-pointing Hamamatsu R10789 series photomultiplier tubes (PMTs)
arranged on an 80\,cm diameter pentakis-dodecahedron support structure within the vessel to give a total 
photocathode coverage of 62.4\,\%. 
These PMTs view an active target region containing 835\,kg of liquid xenon.
In order to  monitor the PMT stability and measure the trigger efficiency,
eight blue LEDs with Teflon diffusers are mounted to the support structure. 
There are six LEDs arranged along the equator and one each at the top and the bottom of the pentakis-dodecahedron.
To shield the scintillator volume from external gammas, neutrons, and muon-induced backgrounds, 
the copper vessel is placed at the center of a $\phi$\,10\,m $\times$ 11\,m 
cylindrical tank filled with pure water.  
This volume is viewed by 72 Hamamatsu R3600 20-inch PMTs to provide both an active muon veto and passive shielding against these backgrounds. 
This is the first water Cherenkov shield used in a dark matter search experiment.
To perform energy and position reconstruction calibrations a portal has been prepared along 
the central vertical axis ($z$-axis) of the PMT support structure through which an OFHC copper rod 
can be inserted directly into the target volume. Thin cylindrical calibration sources  
containing either of $^{55}$Fe, $^{57}$Co, $^{109}$Cd, or $^{241}$Am are placed at the tip of this rod to perform detector calibrations. 
A more detailed description of the XMASS detector is presented in~\cite{xmass_det}.

The data used for this analysis, corresponding to 6.70 days of livetime, was taken in February 2012 with 
a low trigger threshold of four PMT hits~\cite{xmass_lmw}. 
Using 122\,keV gammas from the $^{57}$Co calibration source the xenon light yield 
was found to be 14.7\,photoelectrons/keVee. 
This large light yield allows the analysis threshold to be lowered sufficiently 
for sensitivity to low mass WIMPs.
In order to achieve optimal sensitivity, the entire detector volume is used
because fiducialization is increasingly difficult at these low energies.
A sequence of data reduction is applied to remove events caused by the tail of the scintillation light distribution
after energetic events; (1) events triggered only with the liquid xenon detector are selected,
(2) events that occurred within 10\,ms of the previous event are rejected, and (3) events whose
timing distribution has an RMS greater than 100\,ns are removed. The last cut is applied to remove
Cherenkov events originated from $^{40}$K contamination in the PMT photocathodes; events
with more than 60\% of their PMT hits occurring within the first 20\,ns of the event window are
removed as Cherenkov-like. 

\begin{figure}
\begin{center}
\includegraphics[width=8.5cm]{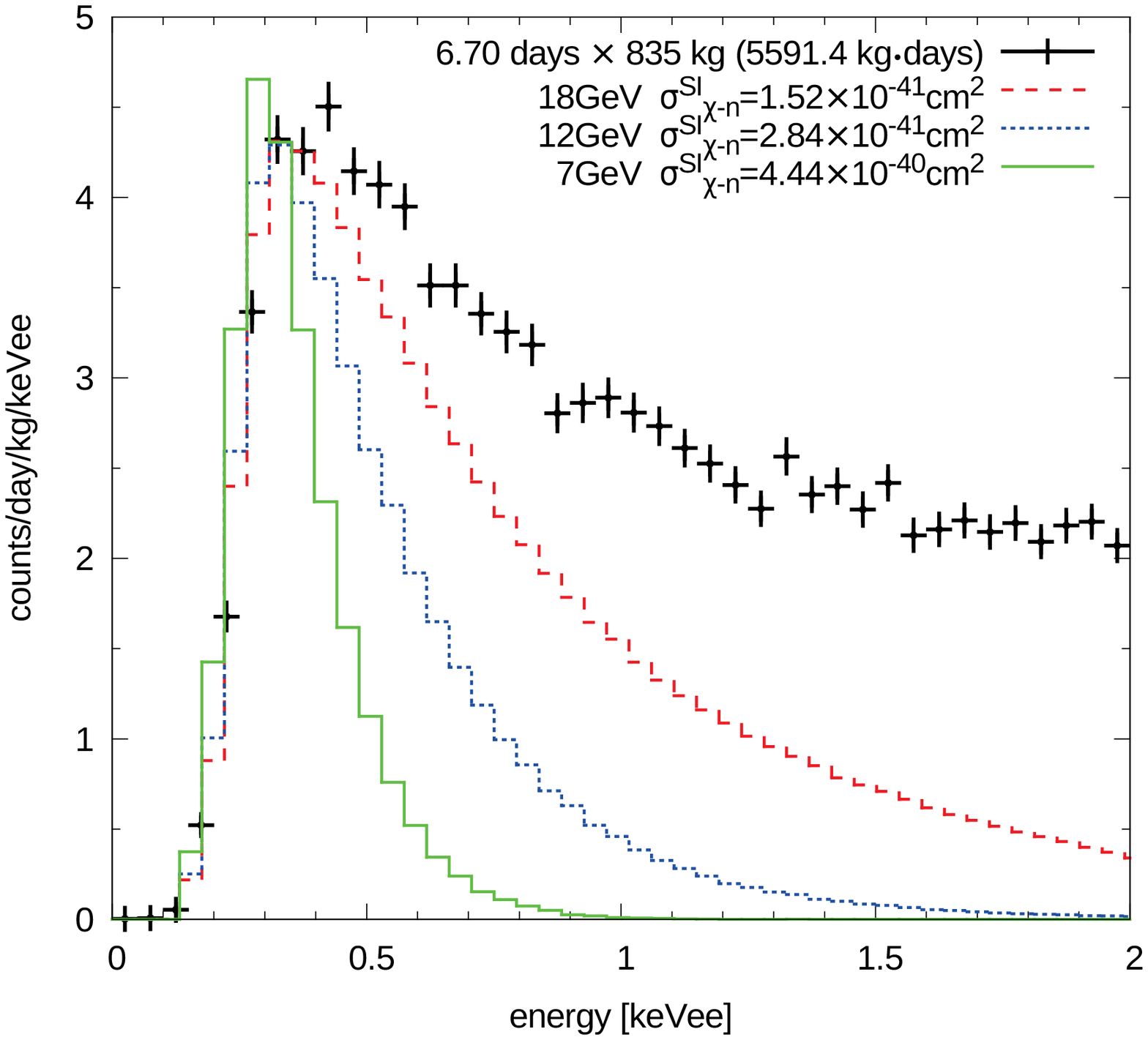}
\caption{Simulated WIMP energy spectra in the XMASS detector assuming the maximum cross section
that provides a signal rate no larger than the observation in any bin above 0.3\,keVee.}
\label{fig:spe}
\end{center}
\end{figure}

\begin{figure}
\begin{center}
\includegraphics[width=8cm, angle=270]{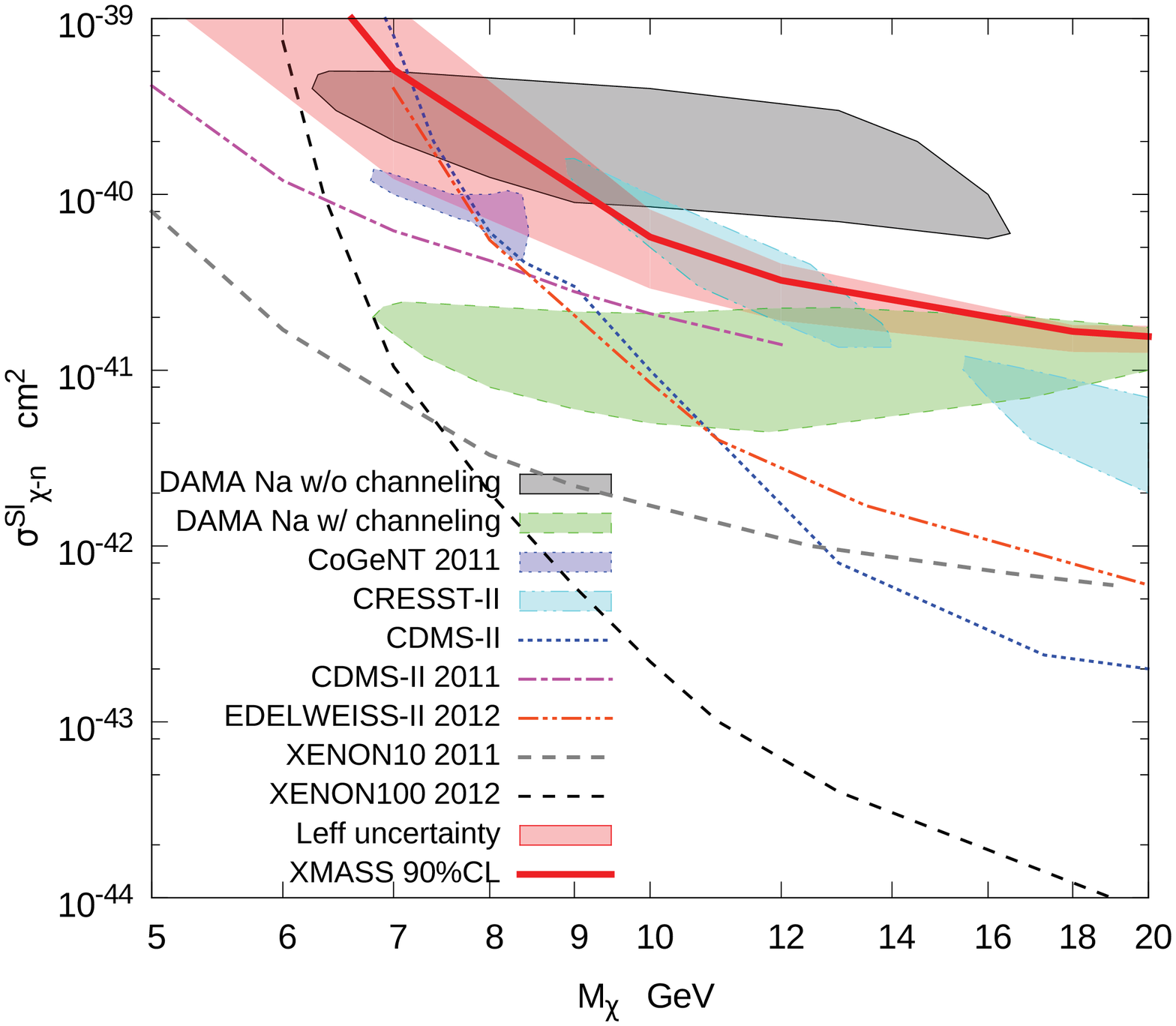}
\caption{Spin-independent elastic WIMP-nucleon cross section as a function of WIMP mass. 
         All systematic uncertainties except that from $\mathcal{L}_{\rm eff}$ are taken into account
         in the XMASS 90\,\% C.L. limit line. The effect of the $\mathcal{L}_{\rm eff}$ uncertainty 
         on the limit is shown in the band.
         Limits from other experiments and favored regions are also shown~\cite{dama, cogent, cresst, xenon, cdms, edelweiss, xenon10}.} 
\label{fig:limit}
\end{center}
\end{figure}

Figure~\ref{fig:spe} shows simulated WIMPs energy spectra overlaid on the observed spectrum 
after the data reduction was applied. 
WIMPs are assumed to be distributed in an isothermal halo with $v_o=220$\,km/s, 
a galactic escape velocity of $v_{\rm esc}=650$\,km/s, and an average density of 0.3\,GeV/cm$^3$. 
In order to set a conservative upper bound on the spin-independent WIMP-nucleon cross section, 
the cross section is adjusted until the expected event rate in XMASS does not exceed the observed one 
in any energy bin above the analysis threshold. It is
chosen as the energy at which the trigger efficiency is greater than 50\% for \,5GeV WIMPs
and corresponds to 0.3\,keVee.
The resulting 90\,\% confidence level (C.L.) limit is shown in Figure~\ref{fig:limit}. 
The impact of the uncertainty from $\mathcal{L}_{\rm eff}$~\cite{Leff} is large in this analysis, so 
its effect on the limit is shown separately in the figure. 

After careful study of the events surviving the analysis cuts, their origins are not completely
understood. Contamination of $^{14}$C in the GORE-TEX$^{\mbox{\scriptsize{\textcircled{\tiny R}}}}$ sheets 
between the PMTs and the support structure may explain a fraction of the events.
Light leaks through this material are also suspect. 
Nonetheless, the possible existence of a WIMP signal hidden under these and other backgrounds cannot be excluded. 
Although no discrimination has been made between nuclear-recoil and electronic events, 
and many events remain in the analysis sample, 
the present result excludes part of the parameter space favored by other 
measurements~\cite{dama,cogent,cresst} when those data are interpreted as a signal for light mass WIMPs.
 
The fiducial volume analysis is ongiong for more sensitive searches.
In addition, we are working on modifications to the inner surface of XMASS, 
especially around the PMTs, to improve the detector performance.
Finally, a R\&D project aiming at modification to single phase liquid xenon TPC with charge readout has been started.

\end{document}